\documentclass[
10pt,
a4paper,
onecolumn
]{article}

\usepackage[margin=3.5cm]{geometry}

\usepackage[utf8]{inputenc}
\usepackage[english]{babel}
\usepackage{csquotes}
\usepackage[T1]{fontenc}
\usepackage{amsmath}
\usepackage{amsfonts}
\usepackage{amssymb}
\usepackage{graphicx}
\usepackage{lmodern}
\usepackage[separate-uncertainty=true,multi-part-units=single]{siunitx}
\DeclareSIUnit\Stokes{St}

\usepackage[style=nature,
maxcitenames=1,
sorting=none,
natbib=true,
backend=biber,
doi=false,
isbn=false,
url=false,
language=english,
]{biblatex}
\bibliography{bib}
\usepackage{hyperref}
\hypersetup{hidelinks}

\usepackage{booktabs}
\usepackage{multirow}
\usepackage{tabularx}
\newcolumntype{Y}{>{\centering\arraybackslash}X}

\usepackage[labelfont=bf]{caption}
\usepackage{float}
\usepackage{subcaption}
\usepackage{rotating}
\graphicspath{{Images/}}

\usepackage{authblk}

\usepackage{xcolor}

\author[1]{Gaby Launay}
\author[2]{Muhammad Subkhi Sadullah}
\author[1]{Glen McHale}
\author[1]{Rodrigo Ledesma-Aguilar}
\author[2]{Halim Kusumaatmaja}
\author[1,*]{Gary G. Wells}
\affil[1]{Smart Materials \& Surfaces Laboratory, Faculty of Engineering \& Environment, Northumbria University, Newcastle upon Tyne, NE1 8ST, UK}
\affil[2]{Department of Physics, Durham University, Durham, DH1 3LE, UK}
\affil[*]{email: gary.wells@northumbria.ac.uk}
\title{Self-Propelled Droplet Transport on Shaped-Liquid Surfaces }
\date{\today}
\begin{document}
\maketitle

The transport of small quantities of liquid on a solid surface is inhibited by the resistance to motion caused by the contact between the liquid and the solid.
To overcome such resistance, motion can be externally driven through gradients in electric fields~\citep{pollack2000electrowetting, jones2001dielectrophoretic}, temperature~\citep{brochard1989motions, darhuber2003microfluidic, yarin2002motion}, light~\citep{oh2002photocontrol, baigl2012photo} and pressure~\citep{squires2005microfluidics}, or structural topography combined with vibration or phase change~\citep{linke2006self, lagubeau2011leidenfrost}, but these all inconveniently involve the input of external energy.
Alternatively, gradients in physical shape and wettability – the conical shape of cactus spines~\citep{ju2012multi} and the wettability of butterfly wings~\citep{liu2014asymmetric} – occur naturally and can be engineered into surfaces~\citep{baigl2012photo, chaudhury1992make, li2018spontaneous, tan2016patterned, bain1994rapid, li2016oil}  to create self-propelled motion.
However, such self-propelled motion to date has limited success in overcoming the inherent resistance to motion of the liquid contact with the solid.
Here we propose a simple solution in the form of shaped-liquid surface, where solid topographic structures at one length scale provides the base for a smaller length-scale liquid conformal layer.
This dual-length scale render possible slippery surfaces with superhydrophobic properties~\citep{dong2018superoleophobic}.
Combined to an heterogeneous topography, it provides a gradient in liquid-on-liquid wettability with minimal resistance to motion and long range directional self-propelled droplet transport.
Moreover, the liquid-liquid contact enables impacting droplets to be captured and transported, even when the substrate is inverted.
These design principles are highly beneficial for droplet transport in microfluidics, self-cleaning surfaces, fog harvesting and in heat transfer.

An ideal design of a system to move small quantities of fluid should not require complex propulsion mechanisms based on continual input of energy.
It should neither present large threshold sticking forces to be overcome nor limit the distance of transport.
The ability to provide fine control of speed of motion, to adhere and transport liquids without boundary walls and to do so in multiple orientations, whether uphill or inverted, would provide additional benefits.
Motion without energy input, has led to a focus on surfaces with a gradient in physical properties and in the use of topographic features~\citep{li2018spontaneous}.
These features provide a gradient in the wetting properties of the solid surface, whilst retaining a uniform surface chemistry, and create a self-propulsion force on droplets.
However, direct contact between the droplet and the solid is needed to drive motion, and this introduces static and dynamic friction forces.
The consequent minimum self-propulsion forces required for transport necessitate large gradients in the topographic texture.
Large gradients, in turn, lead to limited control of a droplet’s velocity when transported over long distances.
To overcome these limitations, we hypothesized that contact with the solid could be completely replaced by another liquid, but in a manner that still allows a gradient in wettability to exist.
To do so, we used a dual length scale substrate providing both a liquid surface and a gradient in its surface texture.
This enables liquid-on-liquid wetting to drive the motion of droplets, where the driving mechanism is mediated by an asymmetry built within an underlying solid substrate which shapes the liquid surface.
To illustrate liquid-on-liquid wetting, first consider placing a water droplet on a thin film of silicone oil that coats a flat hydrophobic solid surface.
Because silicone oil completely wets water, the oil will encase the droplet and isolate it from the solid surface~\citep{smith2013droplet, daniel2017oleoplaning}.

\begin{figure}[p]
  \centering
  \includegraphics[width=.75\textwidth]{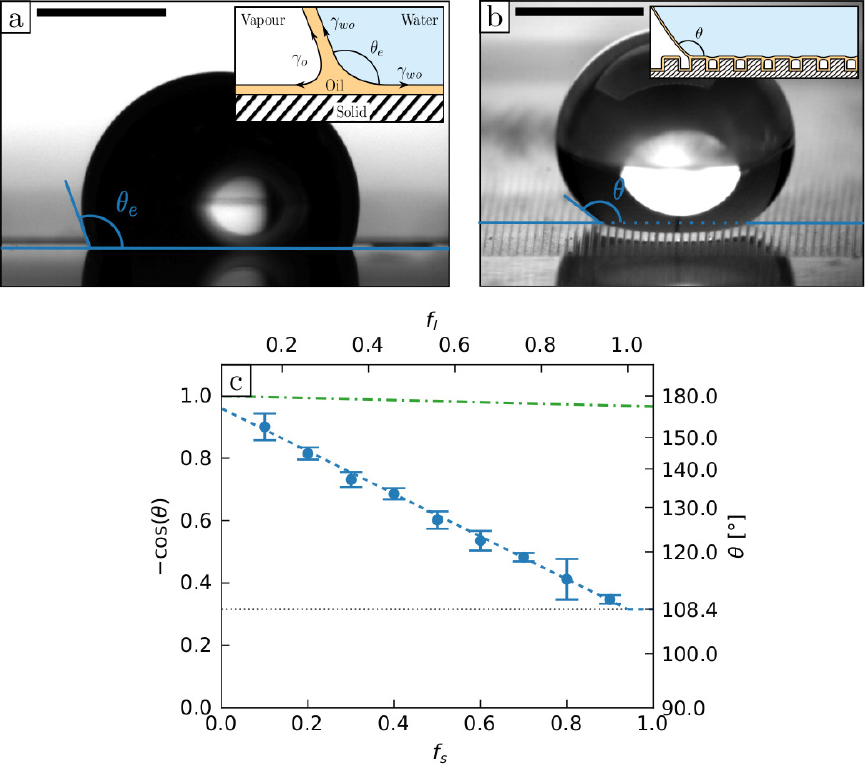}
  \caption{
    \textbf{Wetting on liquid surfaces.}
    (a) Water droplet on a low-hysteresis liquid surface obtained by imbibing an hydrophobic nano-particles coating with silicon oil.
    The scale bar is \SI{1}{\mm}.
    (inset) Diagram presenting the surface tensions acting at the droplet edges.
    (b) Water droplet on a liquid surface shaped with rectangular rails.
    The fraction of the sample covered by the rails is $f_s=0.18$.
    Pockets of air are trapped underneath the droplet, modifying its apparent contact angle.
    The scale bar is \SI{1}{\mm}.
    (inset) Illustration of the expected oil distribution.
    (c) Droplet apparent contact angle, $\theta$, as a function of the rail fraction $f_s$, for \SI{5}{\micro \liter} water droplets deposited on shaped liquid surfaces.
    Error bars are the standard deviation over eight measures.
    (Blue) dashed line is equation~\ref{eq:2} with $f_l = f_s + 0.059$.
    (Green) dashed-dotted line is the Cassie-Baxter equation for a textured oil-free super-hydrophobic surface.
  }
  \label{fig:Figure1}
\end{figure}
Even though there is no direct contact of the droplet with the solid surface, it still exhibits an apparent contact angle, $\theta_e$, due to the interfacial tensions pulling on the outer and inner surfaces of the oil layer at the droplet’s edge (see inset to Fig.~\ref{fig:Figure1}a).
This leads to an effective Young’s Law for liquid-on-liquid wetting~\citep{semprebon2017apparent, kreder2018film}
\begin{equation}
  \label{eq:1}
  \cos\theta_e = \frac{\gamma_{oa} - \gamma_{wo}}{\gamma_{eff}}
\end{equation}
where $\gamma_{eff} = \gamma_{wo} + \gamma_{oa}$. Here, $\gamma_{oa} = \SI{19.8}{\milli \N \per \m}$ and $\gamma_{wo} = \SI{38}{\milli \N \per \m}$ are the surface tensions of the oil-air and water-oil interfaces and predict $\theta_e = \SI{108.4}{\degree}$.
If the oil is replaced by another immiscible liquid having a negative spreading power on water, the liquid may still isolate the droplet from the solid surface but does not encase the droplet and so $\gamma_{eff} = \gamma_{wa}$.
Since equation~\ref{eq:1} has no explicit dependence on the solid surface, it can be regarded as a definition of wettability for an immiscible liquid surface in the thin film limit.
To create liquid-on-liquid wetting with a surface capable of being shaped, we introduce a small length scale solid texture using a hydrophobic nanoparticle-based coating.
This allows the silicone oil to be retained and provides a continuous liquid surface~\citep{wong2011bioinspired, luo2017slippery}.
The apparent contact angle of a droplet of water on this liquid surface is $\theta_e = \SI{109.3 \pm 0.7}{\degree}$ consistent with equation~\ref{eq:1} (Fig.~\ref{fig:Figure1}a) and confirms the absence of direct contact with the solid surface.
Measurements show that the droplet on the liquid surface exhibits a very low contact angle hysteresis ($\Delta \theta \approx \SI{1}{\degree}$) and sliding angle ($\theta_s <\SI{1}{\degree}$ for a $\SI{5}{\micro \liter}$ droplet) thus removing the principal constraint on self-propulsion using a gradient surface.
This approach allows us to create a conformal liquid coating on a solid substrate which can be textured at a larger length scale and so unlocks the ability to create wetting contrasts and gradient wettability with a liquid surface.
We now consider the static wetting of a water droplet on liquid surfaces with textures designed to control wettability.
For the larger length scale, we used arrays of rectangular cross-section rails $\SI{60}{\micro \meter}$ high and $\SI{75}{\micro \meter}$ apart, and with solid surface area fractions $f_s$ between $0.1$ and $0.9$.
 We applied the nanoparticle-based coating as our second, smaller, length scale to enable the silicone oil to create a liquid surface that conforms to the large-scale solid texture.
We observed that a water droplet rests on top of the rails leaving pockets of air underneath~\citep{dai2015slippery} (Fig.~\ref{fig:Figure1}b and inset), with low contact angle hysteresis ($\Delta \theta \approx \SI{1}{\degree}$) and a contact angle (viewed across the rails) which increases with decreasing rail fraction (Fig.~\ref{fig:Figure1}c).
This is reminiscent of the Cassie-Baxter state for superhydrophobic surfaces~\citep{cassie1944wettability, quere2008wetting}, but here the contact angle, $\theta$, is determined by the liquid (equation~\ref{eq:1})~\citep{mchale2019apparent}
\begin{equation}
  \label{eq:2}
  \cos\theta = f_1 \cos\theta_e - (1 - f_1)
\end{equation}
and $f_l \approx f_s$ is the liquid surface area fraction.
This prediction agrees with measurements and provides an excellent fit by including a small correction $f_l=f_s+0.059$ (Fig.~\ref{fig:Figure1}c).
This correction is consistent with the estimated nanoparticle-based coating thickness in our experiments ($\approx \SI{2}{\micro \meter}$, giving a correction of $\approx 0.053$).
This suggests the droplet is in a mixed wetting state suspended above a composite liquid and air surface.
To further confirm there is no contact with the solid we considered the same substrate without silicone oil.
Crucially, because of the high contact angle of $\SI{165}{\degree}$ on this oil-free surface compared to the lower apparent contact angle of $\SI{108.4}{\degree}$ on those with oil, the variation of contact angle with rail fraction is significantly weaker (compare dashed and dashed-dotted lines in Fig.~\ref{fig:Figure1}c).
This implies that the gradient in wettability for droplet self-propulsion is significantly stronger for the liquid surface and amplifies the effect of the liquid surface in removing pinning.
To verify the droplet recognizes contrasts in wettability on a composite liquid-air surface, we placed a droplet at the boundary between two regions of different rail fraction.
The droplet spontaneously moved to the region of higher wettability defined by equation (2) (Supplementary Fig.~S5 and movie~S1).

We now consider the self-propulsion of a water droplet on a liquid surface induced by a designed gradient in wettability.
For the larger length scale, we fabricated an array of $\SI{2}{\cm}$ long and $\SI{60}{\micro \meter}$ high rails, where the width of each rail diverges along its length.
At the low surface fraction end of the substrate, the apex of each rail was $\SI{75}{\micro \meter}$ apart (Supplementary Fig.~S3).
The liquid surface area fractions $f_l$ increases linearly from 0 to 1 at a rate $\alpha = \SI{0.05}{\per \mm}$ as the structure is traversed from the apex to the base.
Without the liquid surface provided by silicone oil, a droplet remains pinned to the solid despite the wettability gradient (Supplementary Fig.~S6).
\begin{figure}[p]
  \centering
  \includegraphics[width=0.75\textwidth]{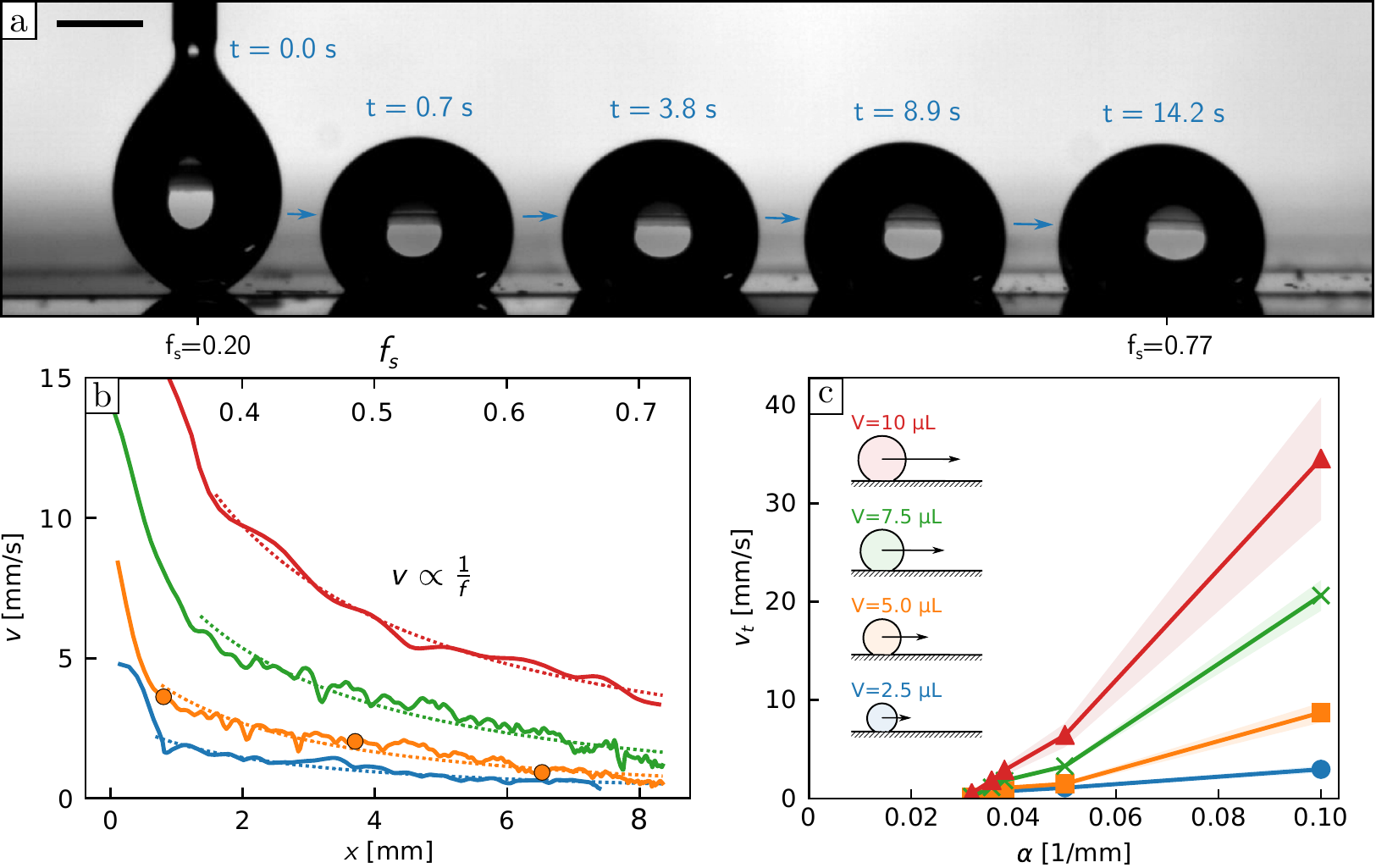}
  \caption{
    \textbf{Self-propulsion on a textured gradient liquid surface.}
    (a): \SI{5}{\micro \liter} water droplet propelled on a textured liquid surface.
    The gradient of rail fraction ($\alpha =   \mathrm{d} f_s / \mathrm{d} x = \SI{0.05}{\per \mm}$) induces the wettability gradient driving the droplet.
    The scale bar is \SI{1}{\mm}.
    (b): Droplet speed during motion for different volumes:
    \SI{2.5}{\micro \liter} (blue), \SI{5}{\micro \liter} (yellow), \SI{7.5}{\micro \liter} (green) and \SI{10}{\micro \liter} (red).
    (Yellow) circles correspond to the droplet at three of the times presented in (a) ($t=0.7$\SI{}{\second}, $t=3.8$\SI{}{\second} and $t=8.9$\SI{}{\second}).
    After an initial transient regime due to the deposition of the droplet, the velocity decreases in $1/f$ along the sample (dashed lines as fits).
    (c): Typical speed $v_t$ of water droplets as a function of gradient of rail fraction $\alpha$ and droplet volumes  (blue circles: \SI{2.5}{\micro \liter}, orange squares: \SI{5}{\micro \liter}, green crosses: \SI{7.5}{\micro \liter} and red triangles: \SI{10}{\micro \liter}).
  }
  \label{fig:Figure2}
\end{figure}
On the shaped liquid surface, however, we observed a sustained self-propulsion over the pattern (Fig.~\ref{fig:Figure2}a).
This propulsion can also be designed into complex pathways, as illustrated by transport along a curved path (Supplementary Movie~S3).
Returning to our design criteria and the desire for controlled long distance transport and fine control of droplet speed, we discovered an inverse linear relationship between speed of motion and rail fraction $v \propto 1/f_s$ (Fig.~\ref{fig:Figure2}b).
The slowing-down dynamics occurs for a range of droplet volumes ($V = 2.5 - 10 \si{\micro \liter}$) and wettability gradients ($\alpha = 0.03 - 0.1 \si{\per \mm}$).
By balancing the driving force $F_d \propto \alpha \gamma_{oa} R^2$ due to the wettability gradient along the droplet’s perimeter (where $R$ is the droplet’s base) with viscous resistance $F_v \propto \mu_o f_s R v$ from the edge of the drop, we find $v \propto \gamma_{oa} \alpha R/\mu_o f_s$ (Supplementary Information).
This explains the observed slowing-down with increasing rail fraction (Fig.~\ref{fig:Figure2}b) and predicts an increase of speed with wettability gradient and droplet volume, which we confirmed by measuring the typical droplet speed, $v_t$, defined as the speed at the surface point where $f_s = 0.5$ (Fig.~\ref{fig:Figure2}c).
Our measurements show self-propulsion ceases below a rail fraction gradient $\alpha \approx \SI{0.03}{\per \mm}$ implying the presence of a small pinning force (Fig.~\ref{fig:Figure2}c).
\begin{figure}[p]
  \centering
  \includegraphics[width=.75\textwidth]{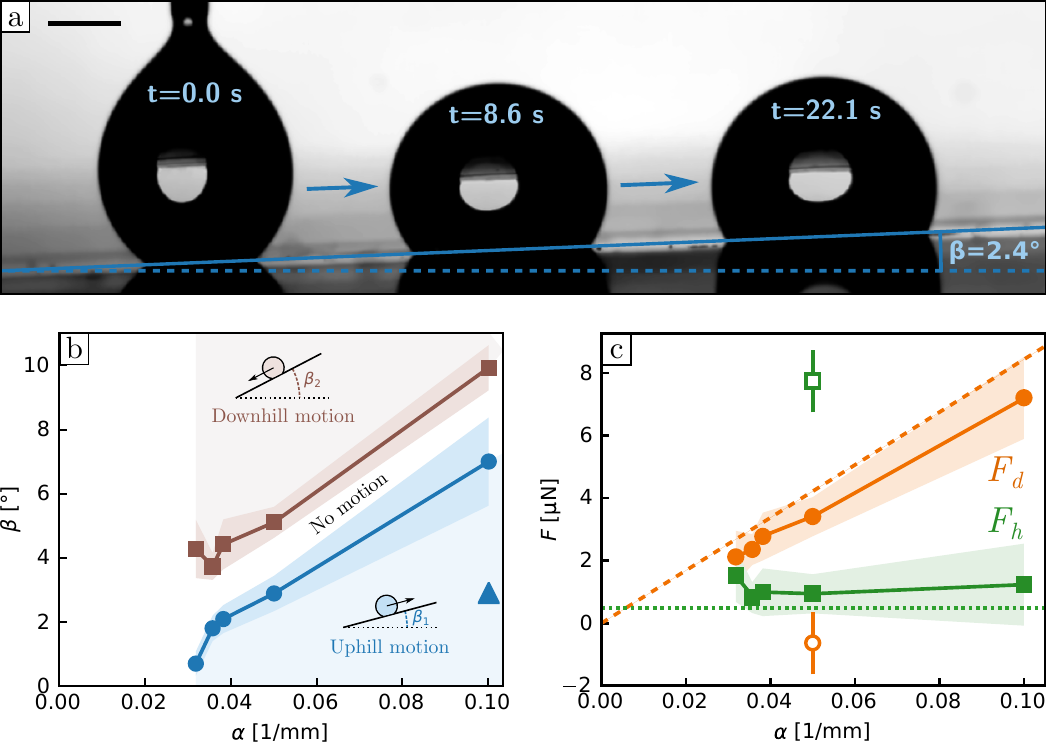}
  \caption{
    \textbf{Strength of gradient induced self-propulsion.}
    (a): \SI{5}{\micro \liter} water droplet propelled uphill on a textured liquid surface tilted at an angle $\beta = 2.4$\SI{}{\degree}.
    The scale bar is \SI{1}{\mm}.
    (b): Critical sliding angles $\beta$ as a function of rail fraction gradient $\alpha$,
    for \SI{5}{\micro \liter} water droplets.
    (blue) circles stand for the critical tilting angles $\beta_1$ at which droplets stop moving uphill and (brown) squares stand for the critical tilting angles $\beta_2$ at which droplets stop moving downhill.
    Dim areas represent the standard deviation calculated on 12 measurements.
    Between those two critical angles, the driving force is too small to overcome the pinning force and the droplet remains motionless.
    The (blue) triangle highlights the configuration depicted in (a).
    (c) Driving force $F_d$ and pinning force $F_p$ calculated from the critical angle measurements presented in (b).
    (orange) circles are the driving force for liquid (filled symbols) and super-hydrophobic (hollow symbol) surfaces.
    (green) squares are the pinning force for liquid (filled symbols) and super-hydrophobic (hollow symbol) surfaces.
    Dim areas represent the standard deviation propagated from the measurement of the critical angles.
    Dashed (orange) and dotted (green) lines are the theoretical predictions of respectively the driving force and the pinning force (see text).
  }
  \label{fig:Figure3}
\end{figure}
To measure this force, we tested the ability of droplets to move up inclined surfaces (Fig.~\ref{fig:Figure3}a).
We observed upwards self-propulsion below an inclination angle $\beta_1$, implying that the driving force outweighs gravity and pinning (bottom region in Fig.~\ref{fig:Figure3}b), but above an angle $\beta_2$ droplets run downhill (upper region in Fig.~\ref{fig:Figure3}b).
From the conditions for mechanical equilibrium, the driving force is $F_d = \frac{1}{2} \rho_d V g ( \sin\beta_2 + \sin \beta_1)$ and the pinning force is $F_p = \frac{1}{2} \rho_d V g ( \sin\beta_2 - \sin \beta_1)$.
We deduce $F_p \approx \SI{1.11 \pm 0.25}{\micro \N}$ across the range of rail fraction gradients, similar to the small value estimated from contact angle hysteresis measurements, $F_p \approx  \SI{0.48}{\micro \N}$ (Supplementary Information).
Fig.~\ref{fig:Figure3}c shows the measured driving force also agrees well with the simple theoretical model (Supplementary Information).
Overall, these measurements confirm that pinning is overcome by the driving force over a broad range of wettability gradients and droplet volumes.
With such small gradients (down to $\alpha \approx \SI{0.03}{\per \mm}$) sustained transport is possible over long distances (here $L = 1/\alpha \approx \SI{30}{\mm}$).
Furthermore, because of the small pinning force, droplets can also be moved at low speeds (down to $\SI{0.067}{\mm \per \second}$ for a $\SI{5}{\micro \liter}$ droplet).
In contrast, without the shaped liquid surface (Fig.~\ref{fig:Figure3}c), we find that the pinning force is considerably larger and cannot be overcome by the driving force.
In addition to reducing pinning and allowing self-propulsion, the liquid surface significantly increases the ability to capture impacting droplets.
The number of bounces of a droplet before it adheres to a solid upon impact depends on the kinetic and surface energy remaining after energy dissipation because of spreading and retraction~\citep{mao1997spread, yarin2006drop}.
\begin{figure}[p]
  \centering
  \includegraphics[width=0.75\textwidth]{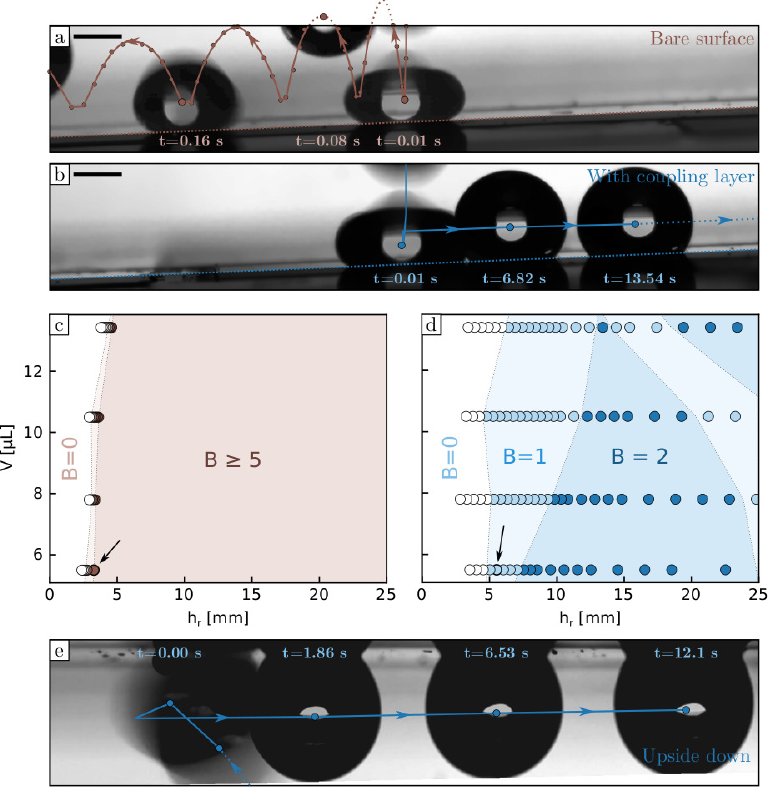}
  \caption{
    \textbf{Impact, capture and transport of droplets.}
    (a): \SI{5}{\micro \liter} water droplet impacting on a tilted ($\beta = 2.33$\SI{}{\degree}) super-hydrophobic textured surface.
    Circles are the successive measured positions of the droplet.
    Due to the low normal adhesion, the droplet bounces and is transported downhill by the effect of gravity.
    (b): Same as (a) but for a liquid surface.
    The higher energy dissipation during impact and the strong wettability gradient allow the droplet to be captured rapidly and transported uphill.
    (c): Number of bounce $B$ as a function of droplet volume and release height on a super-hydrophobic tilted surface.
    The black arrow highlights the configuration depicted in (a).
    % The bouncing rate $B$ increases quickly with the release height, indicating a poor ability to capture droplets.
    (d): Similar as (c) but for a liquid surface.
    For all points on this plot, the droplets are successfully captured and moved uphill.
    The black arrow highlights the configuration depicted in (d).
    (e): Hanging droplet, captured and transported on an inverted shaped liquid surface.
  }
  \label{fig:Figure4}
\end{figure}
On a superhydrophobic surface, the small contact area and low dynamic friction of a droplet leads to a bouncing rate that rapidly increases with increasing droplet impact speed (which we varied by setting the initial release height $h_r$ (Fig.~\ref{fig:Figure4}a,c), making droplet capture and transport difficult.
In contrast, our shaped liquid surfaces increase dissipation upon impact through viscous friction~\citep{kim2016droplet} and have stronger adhesion due to the wettability of the liquid surface.
As such, they prevent bouncing over a broad range of release heights and promote capture and subsequent transport via self-propulsion, even uphill (Fig.~\ref{fig:Figure4}b,d).
Furthermore, Fig.~\ref{fig:Figure4}e shows that these surfaces are capable of capturing and transporting hanging droplets, impacting the surface from below.

The shaping of liquid surfaces, enabled by a dual length-scale structure, is an advantageous approach to creating wettability contrasts for droplets on liquid surfaces. In this work we used these principles to design liquid surfaces with strong gradients in wettability and low resistance to droplet motion. On a shaped liquid surface, it is possible to gently transport droplets with ease over long distances along defined paths and with fine control of their speed. The design principles presented here provide a foundation for new types of microfluidic devices and, because of the ability to capture and transport impacting droplets in any orientation, may also have wider applications, such as fog harvesting and heat transfer.

\printbibliography
% \begin{thebibliography}{}
%   \input{drop_motion_article.bbl}
% \end{thebibliography}

\clearpage
\section{Supplemental Material}

\subsection{Surface fabrication process}
\label{sec:fabrication-process}
\begin{figure}[H]
  \centering
  \includegraphics[width=\textwidth]{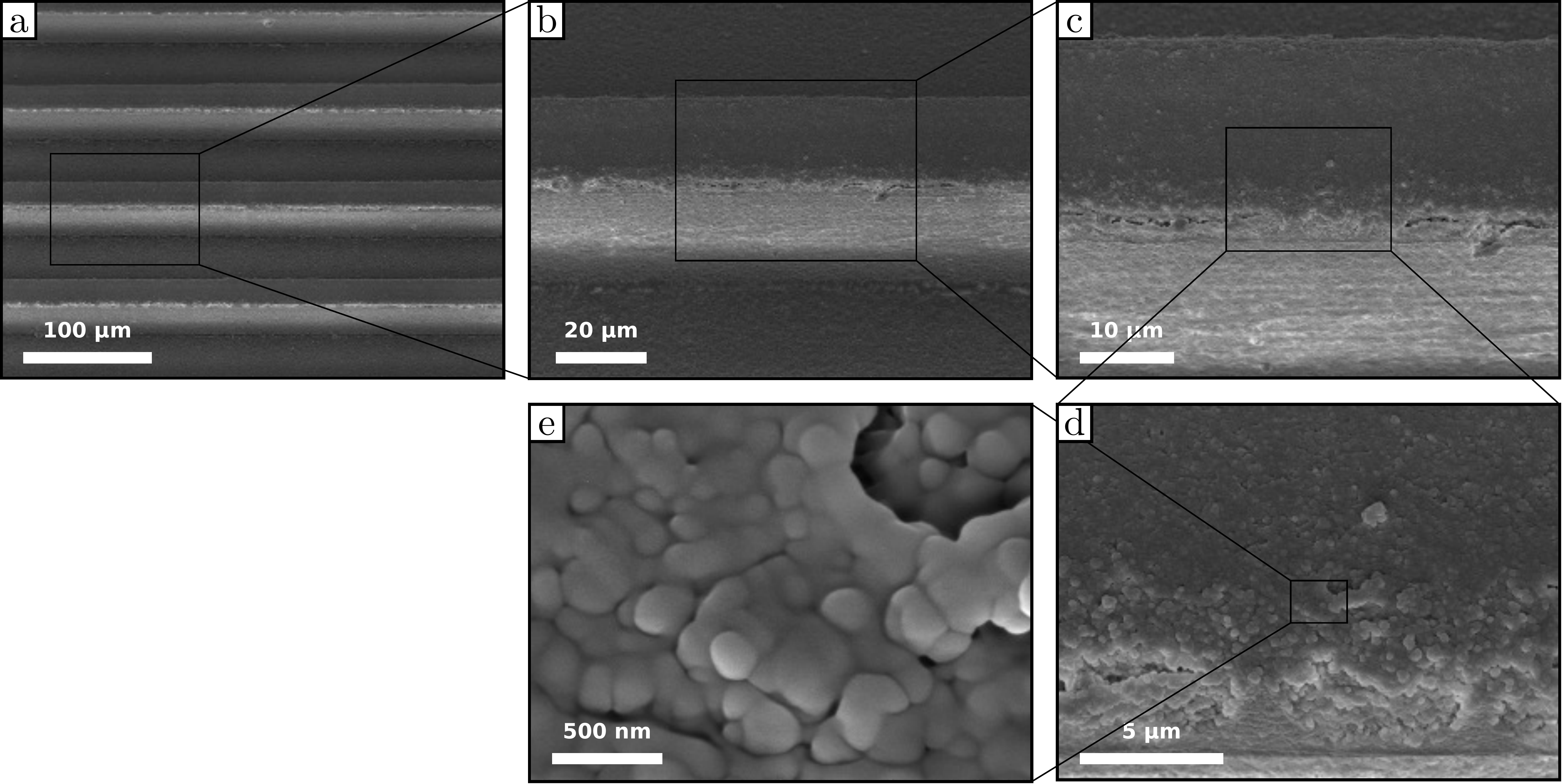}
  \caption{
    SEM images of the micro-structures manufactured by photolithography and coated with a conformal layer of silanized nano-particles.
    (a) Parallel rails obtained by photolithography.
    (e) Close-up of the top of a rail highlighting the nano-particles of the porous layer.
  }
  \label{fig:Fabrication_glaco_layer}
\end{figure}

\begin{figure}[H]
  \centering
  \includegraphics[width=\textwidth]{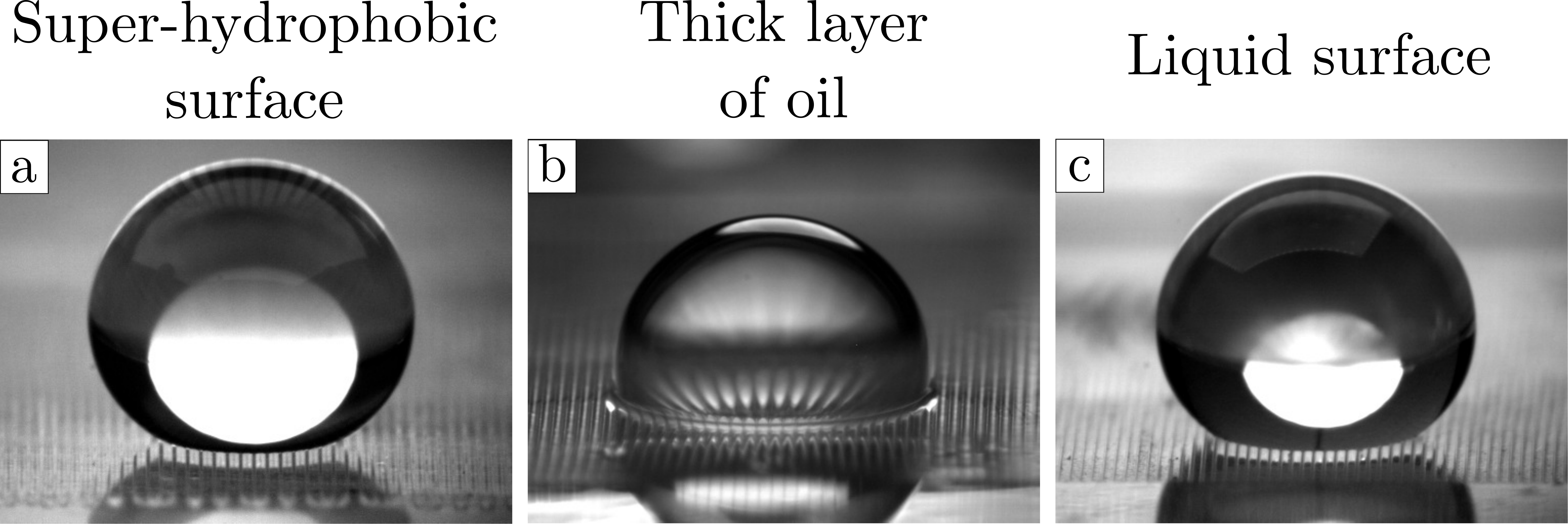}
  \caption{
    \SI{5}{\ul} droplets on patterned surfaces.
    (a) Super-hydrophobic patterned surface.
    The droplet lies in Cassie-Baxter state on the top of rails.
    (b) After dip-coating the surface in silicone oil, the oil completely fills the micro-structures.
    The contact angle hysteresis is then very low and the droplet highly mobile, but the amount of oil prevents the formation of pockets of air underneath the droplet.
    (c) After removing the excessive oil, we obtain a shaped liquid surface, that exhibits a high droplet mobility, while still allowing the formation of air pockets underneath the droplet.
  }
  \label{fig:VisuCB}
\end{figure}

\textbf{We describe here the fabrication process of the super-hydrophobic and shaped liquid surfaces.}\\
Micro-structured surface are manufactured by photolithography.
Glass wafers are cleaned prior to the process, using IPA and acetone rinses.
To promote the photoresist adhesion, wafers are etched using a reactive-ion etcher  (RIE) and bounded with HDMS (Microchem).
Surfaces are then spin-coated \SI{30}{\s} at \SI{1750}{rpm} with a photoresist (SU8-2025 from Microchem), to obtain a layer of \SI{60}{\um}.
After a pre-bake (\SI{10}{\minute} at \SI{95}{\celsius}), patterns are transposed from a chrome mask (JD Photo Data) on the surface using a mask aligner (at an UV power of $\SI{300}{\milli \J \per \cm^2}$).
The extra photoresist is removed using EC solvent (Microchem) after a post-exposure bake (\SI{6}{\minute} at \SI{95}{\celsius}).
Surfaces are then subjected to a last bake (\SI{10}{\minute} at \SI{170}{\celsius}) to finalize the cross-linking process.
The structures obtained are \SI{60}{\um} high and can be as small as \SI{10}{\um} wide.
The hydrophobic porous layer is deposited by spraying a commercial solution (Glaco\textsuperscript{\texttrademark} mirror coat zero\textsuperscript{\textregistered}) of silanized nano-particles.
SEM images show that this process creates a conformal porous layer on the micro-structures (Fig.~\ref{fig:Fabrication_glaco_layer}).
At this point, the surface is a super-hydrophobic patterned surface, on which water droplets lie in Cassie-Baxter state (Fig.~\ref{fig:VisuCB}a).
\SI{20}{\centi\Stokes} silicone oil is imbibed into the surface by dip-coating it at \SI{0.1}{\mm \per \s}.
This result in a thick oil layer that fills the micro-structures and prevents the formation of air pockets underneath the droplet (Fig.~\ref{fig:VisuCB}b).
To obtain a liquid surface, the excess oil is removed using a water jet.
The success of this operation is ensured by verifying that deposited droplets are in Cassie-Baxter state (Fig.~\ref{fig:VisuCB}c) and exhibit a very low contact angle hysteresis.

\subsection{Micro-structure geometry}
\begin{figure}[H]
  \centering
  \includegraphics[width=.65\textwidth]{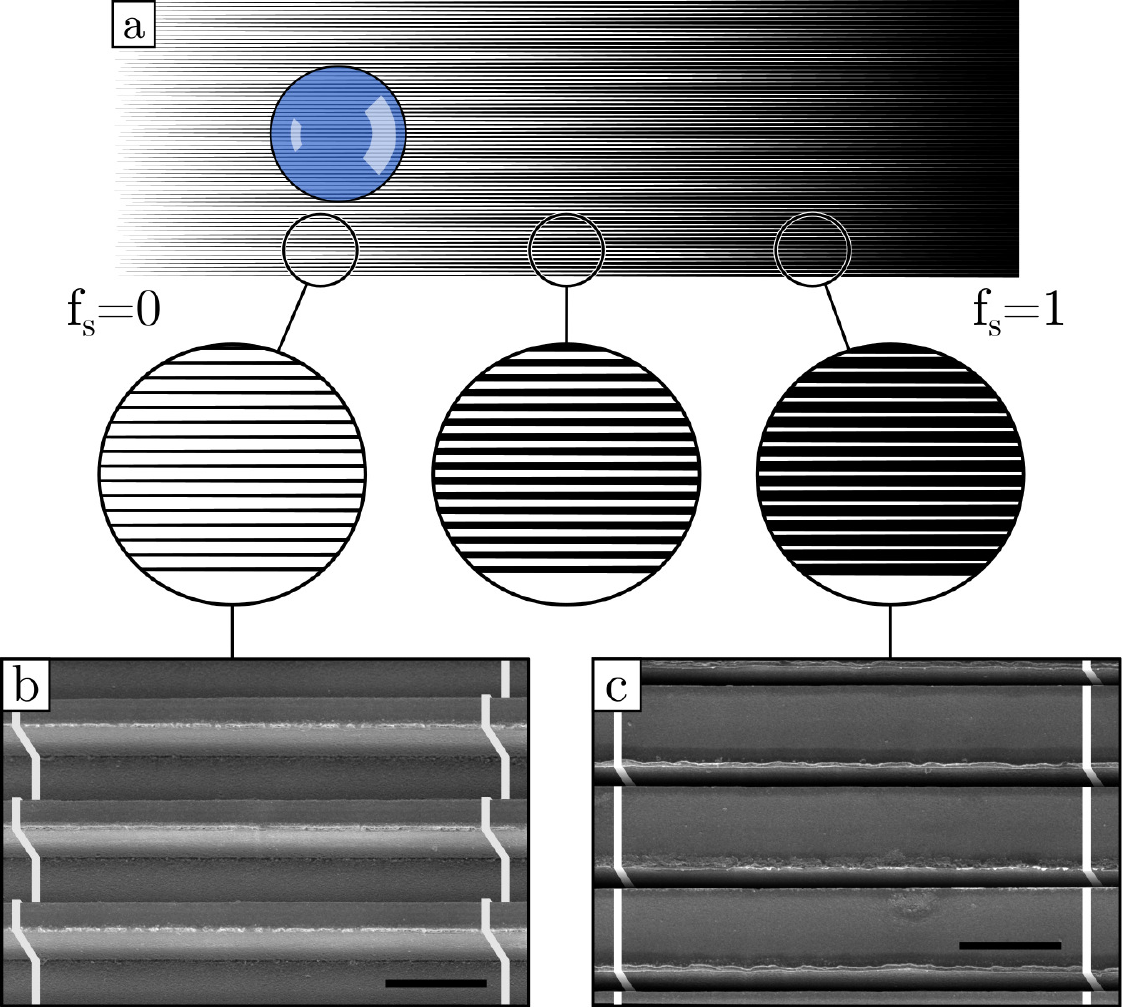}
  \caption{
    Rail geometry used to create gradients of wettability on super-hydrophobic and liquid surfaces.
    (a) Parallel rails with divergent width along their length.
    The center-lines of two consecutive rails are separated by \SI{75}{\um}.
    The solid surface area evolves from $f_s=0$ on the left to $f_s=1$ on the right.
    (b)-(c) SEM images of the divergent rails at two different positions.
    The scale bars are \SI{75}{\um}.
  }
  \label{fig:rail_geometry}
\end{figure}

\textbf{Fig. \ref{fig:rail_geometry} presents the rail geometry used to create wettability gradients with liquid surfaces.}\\

\subsection{Contact angle measurements}
\textbf{We present here the methods used to measure droplet contact angles and contact angle hysteresis.}\\
Static contact angles are measured using side visualizations of droplets and an in-house drop shape analyzer software (\href{https://framagit.org/gabylaunay/pyDSA}{PyDSA}).
The robustness of the measurements is ensured by performing repeats and by comparing the different methods of contact angle measurements provided by the software.

The contact angle hysteresis is measured using the inflation-deflation method.
A droplet of \SI{4}{\uL} is first deposited on the surface and put in contact with a small syringe needle.
The side-view of the droplet is recorded using a camera while the drop is inflated and then deflated slowly through the needle.
The evolution of the base radius and contact angles are extracted from the resulting video using \href{https://framagit.org/gabylaunay/pyDSAqt5}{pyDSA}.
From those measurements, the advancing and receding contact angles are deduced as the contact angles for which the base radius begins to respectively increase and decrease.
The contact angle hysteresis is then calculated as the difference between the advancing and receding angles.
For a liquid surface, measurements indicate that the CAH remains low ($\Delta\theta = \SI{0.66 \pm 0.57}{\degree}$ on 20 measurements) and is independent of the solid fraction in the range $f_s \in [0.1, 0.9]$.

%%% Old description + figure for when the CAH was evolving with f %%%
% \subsection{Contact angle hysteresis measurements}
% Figure~\ref{fig:CAH} presents measurements of the contact angle hysteresis.
% The data scattering is a direct consequence of (i) the difficulty to measure small contact angle hystereses and (ii) small local defects of the surfaces.
% However, the contact angle hysteresis is at most \SI{8}{\degree}, which is sufficiently low to ensure high droplet mobility.
% \begin{figure}[h!]
%   \centering
%   \includegraphics[width=.75\textwidth]{CAH.pdf}
%   \caption{
%   Contact angle hysteresis ($\Delta \theta$) measurements for increasing rail fractions $f$.
%   The contact angle hysteresis remains small $<\SI{7}{\degree}$ for any rail fraction.
%   Despite the scattering, a decrease of the contact angle hysteresis with the rail fraction is clearly visible.
%   This evolution can be explained by the dependence of the contact angle hysteresis to the static contact angle value.
%   The contact angle hysteresis for a rail fraction of $f=0.5$ is estimated using a linear interpolation and read $\Delta \theta (f=0.5) = \SI{3.94}{\degree}$.
% }
%   \label{fig:CAH}
% \end{figure}

\subsection{Apparent contact angle evolution with the solid fraction}
\begin{figure}[H]
  \centering
  \includegraphics[width=.75\textwidth]{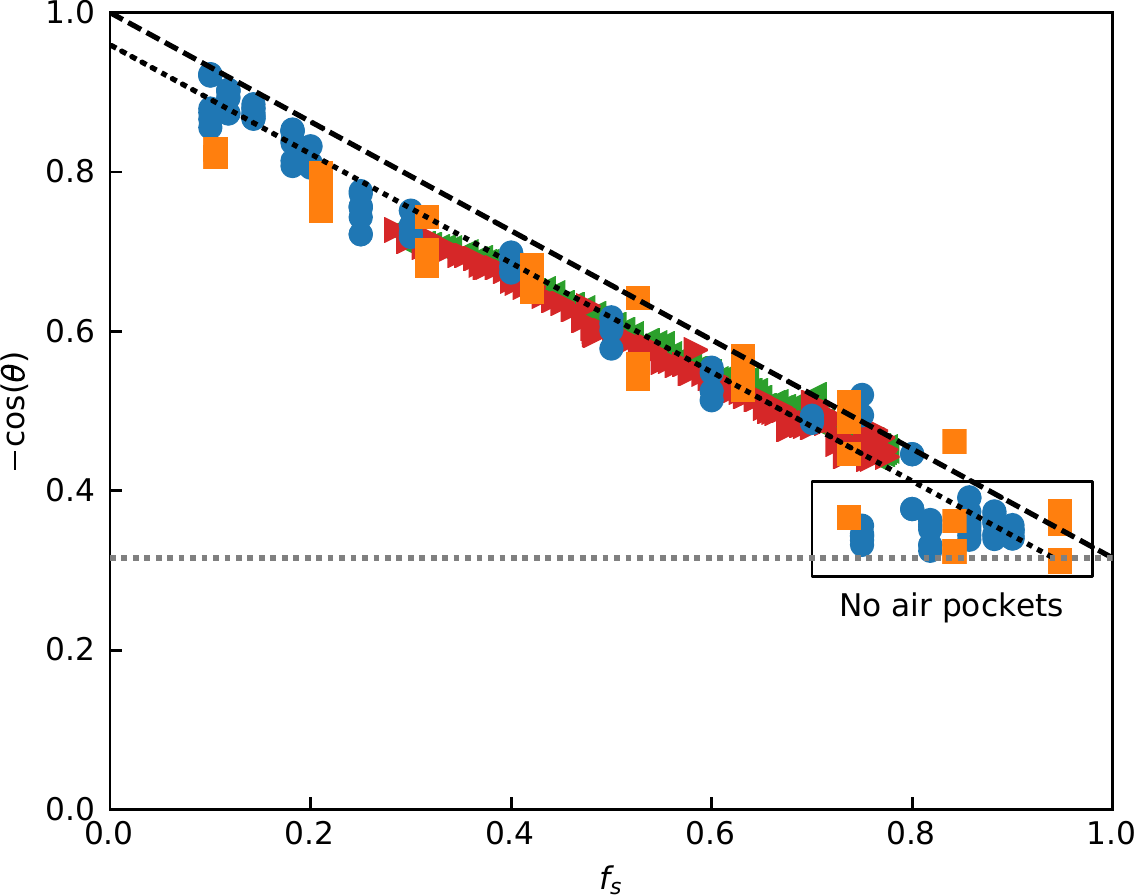}
  \caption{
    Droplet apparent contact angle as a function of the solid fraction $f_s$.
    (blue) circles: static droplets deposited on rails with an uniform solid fraction.
    (orange) squares: droplets pinned in place at different positions on rails with divergent width.
    (red) right-pointing triangles and (green) left-pointing triangles: respectively left and right contact angle of a droplet moving on rails with divergent width.
    Dashed line: Equation (2) of the main manuscript, using the solid fraction $f_s$ and $\theta_e=\SI{108.4}{\degree}$.
    Dotted line: Equation (2) of the main manuscript, using the corrected liquid fraction $f_l=f_s + 0.059$.
    The black box highlights the configuration where air pockets underneath the droplet are filled with oil.
  }
  \label{fig:CA_with_frac}
\end{figure}

\textbf{We present here additional measurements of the droplet apparent contact angle as a function of the solid fraction $f_s$.}\\
Fig. 1c of the main manuscript shows that the apparent contact angle depends linearly of the solid fraction $f_s$ for a droplet on rails with uniform fraction.
Fig.~\ref{fig:CA_with_frac} indicates that this is also the case on tails with divergent width, for both pinned and moving droplets.
For high solid fractions ($f_s > 0.7$), the pocket of air expected to be sustained underneath the droplet are sometimes filled with oil.
This explains the decrease of the apparent contact angle to $\theta_e$, as predicted by Equation 1 of the main manuscript, for some measurements in this region.

\subsection{Droplet response to a wettability contrast on a shaped liquid surface}
\label{sec:dropl-react-wett}
\begin{figure}[H]
  \centering
  \includegraphics[width=.75\textwidth]{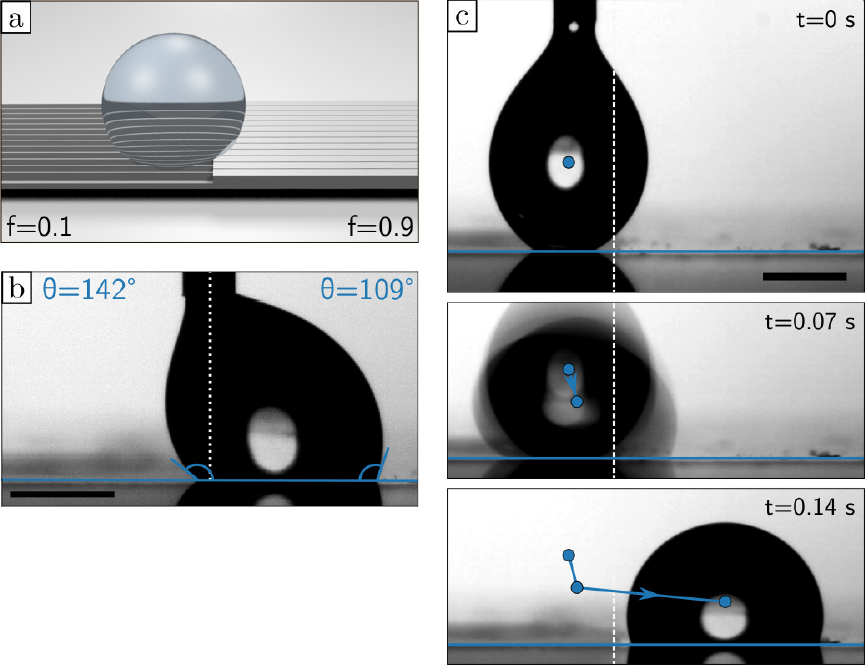}
  \caption{
    Droplet rapid motion due to a discontinuity in wettability of a shaped liquid surface.
    (a) Illustration of a droplet on a discontinuity of solid fraction.
    The solid fraction is $f_s=0.1$ on the left and $f_s=0.9$ on the right.
    (b) Droplet held in place on the discontinuity to highlight the difference in wettability between the two sides.
    (c) Droplet released on the discontinuity and moving rapidly towards the region of higher wettability (higher solid fraction).
    The scale bars are \SI{1}{\mm}.
  }
  \label{fig:wettability_contrast}
\end{figure}

\textbf{A droplet deposited between two regions of different rails fraction spontaneously moves towards the region of higher wettability (higher fraction) (Figure \ref{fig:wettability_contrast}).}\\

See also Movie S1.

\subsection{Droplet on patterned super-hydrophobic surfaces}

\begin{figure}[H]
  \centering
  \includegraphics[width=\textwidth]{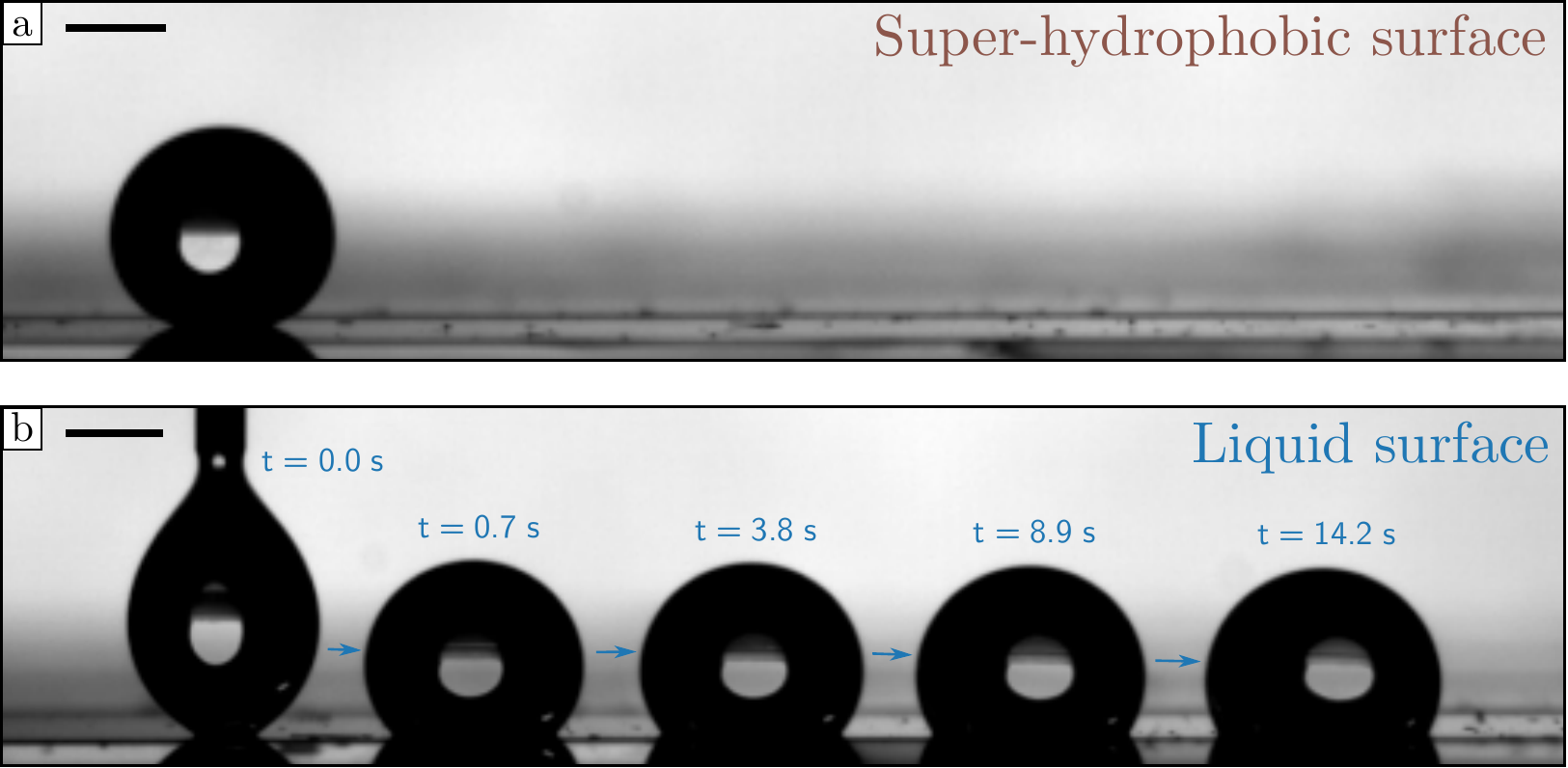}
  \caption{
    Droplets deposited on surfaces with a gradient of solid fraction ($\alpha = \SI{0.05}{\per \mm}$).
    (a) Super-hydrophobic surface.
    The gradient of wettability is not strong enough to overcome the pinning force and the droplet remains motionless.
    (b) Liquid surface.
    The increased gradient of wettability and decreased pinning force allow the droplet to move towards the high wettability regions (\textit{i.e.} lower apparent contact angle as defined by Eq. 1 of the main manuscript).
    The scale bars are \SI{1}{\mm}.
  }
  \label{fig:Gradient_SHS}
\end{figure}

\textbf{We compare here the effect of the wettability gradient on a droplet motion for a super-hydrophobic and a liquid surface (Figure \ref{fig:Gradient_SHS}).}\\

See also Movie S2.

\subsection{Force balance model}
\textbf{We derive here the model used to estimate the driving and pinning forces for a droplet on a shaped liquid-surface.}\\

\subsubsection{Driving force}
The elementary force per unit length acting at a specific point of the contact line of a droplet (and towards the droplet center) reads:
\begin{equation}
  \label{eq:4}
  g = \left( \gamma_{oa} + \gamma_{wo} \right) \cos \theta
\end{equation}
The resulting force acting on the droplet can be obtained by integrating this elementary force on the contact line.
In the case of a droplet on a continuous, linear gradient of solid fraction $\alpha = \frac{\partial f_s}{\partial x}$, and assuming the droplet to have a rectangle footprint of length $R$ and width $kR$, the capillary force acting on the droplet is simply:
\begin{equation}
  F_d = 2 k R \left[ g(x=-R) - g(x=R) \right]
  \label{eq:drop_driving_force}
\end{equation}
Using Eq. 2 from the main manuscript and Eq.~\ref{eq:4}, Eq.~\ref{eq:drop_driving_force} simplifies in
\begin{equation}
  \label{eq:6}
  F_d = 8 k \gamma_{oa} \alpha R^2
\end{equation}

\subsubsection{Viscous resistance}
As soon as a droplet is put in movement, a force due to viscous friction apposes its motion.
Different mechanisms participate to this viscous dissipation: the dissipation in the droplet bulk, in the oil layer underneath the droplet and at the droplet contact line.
The force linked to the dissipation in the droplet bulk $F_{v,b}$ is expected to generate a resisting force scaling as
\begin{equation}
  \label{eq:7}
  F_{v,b} = \int_{S_b} \mu_w \frac{\partial v_b}{\partial y}~dS_b \propto f_s \mu_w v R
\end{equation}
with $S_b$ the surface of contact between the droplet and the surface and $v_b$ the velocity in the droplet bulk.
Assuming the velocity at the top of the oil layer to be $v_o(y=h) = vh\mu_w / R\mu_o$ (with $h$ the oil thickness), the force linked to the dissipation in the layer of oil scales as
\begin{equation}
  \label{eq:7b}
  F_{v,o} = \int_{S_b} \mu_o \frac{\partial v_o}{\partial y}~dS_b \propto f_s \mu_w v R
\end{equation}
Using a Taylor expansion of the Cox-Voinov equation, we can show that the force linked to the dissipation at the contact line scales as:
\begin{equation}
  \label{eq:9}
  F_{v,c} \propto f_s \mu_o v \ln \left( \frac{R}{L_o} \right) R
\end{equation}
with $L_o$ the slip length.
All those forces scale as $f_s v R$.
However, the force linked to the dissipation at the contact line is expected to be dominant in our configuration because of the viscosity ratio ($\mu_o/\mu_w = 20$) and the high ratio $R/L_o$.
Being the dominant dissipation force, $F_{v,c}$ will be called $F_v$ in the sequel.
\subsubsection{Pinning force}
Because of chemical and topological heterogeneities, a droplet on a surface always exhibits a contact angle hysteresis.
This hysteresis leads to the presence of a pinning force, that need to be overcome before a droplet can be put in motion.
Still assuming a droplet with a rectangular footprint, the pinning force can be deduced from the advancing and receding contact angles (respectively $\theta_a$ and $\theta_r$), as
\begin{equation}
  F_p = k R \left( \gamma_{wo} + \gamma_{oa} \right) \left(\cos  \theta_a - \cos  \theta_r \right)
  \label{eq:hysteresis}
\end{equation}
The difference in contact angle cosine can be rewritten as:
\begin{align}
  \cos\theta_a - \cos\theta_r
  =& -2 \sin \left( \frac{\theta_a + \theta_r}{2} \right) \sin \left( \frac{\theta_a - \theta_r}{2} \right) \\
  =& -2 \sin \theta_{eq} \sin \left( \frac{\Delta \theta}{2} \right)
     \label{eq:sin_relation}
\end{align}
with $\Delta \theta = \theta_a - \theta_r$ the contact angle hysteresis
and $\theta_{eq}$ the equilibrium contact angle, estimated to be equal to $(\theta_a + \theta_b)/2$.
Introducing Equation \ref{eq:sin_relation} into \ref{eq:hysteresis} leads to:
\begin{equation}
  F_p = - k R \left( \gamma_{wo} + \gamma_{oa} \right) 2 \sin\theta_{eq} \sin \left( \frac{\Delta \theta}{2} \right)
  \label{eq:hysteresis_final}
\end{equation}
\subsubsection{Terminal velocity}
A droplet terminal velocity can be estimated by balancing the driving and dissipation forces (respectively  $F_d$ and $F_v$):
\begin{equation}
  8 k \gamma_{oa} \alpha R^2 \propto  f_s \mu_o \nu ln \left(\frac{R}{L_o} \right) R
\end{equation}
which can be rewritten as
\begin{equation}
  v \propto \frac{8k\gamma_{oa}}{\mu_o \ln \left( R/L_o \right) } \frac{\alpha R}{f_s}
\end{equation}

\subsection{Hanging droplet motion}
\begin{figure}[H]
  \centering
  \includegraphics[width=.75\textwidth]{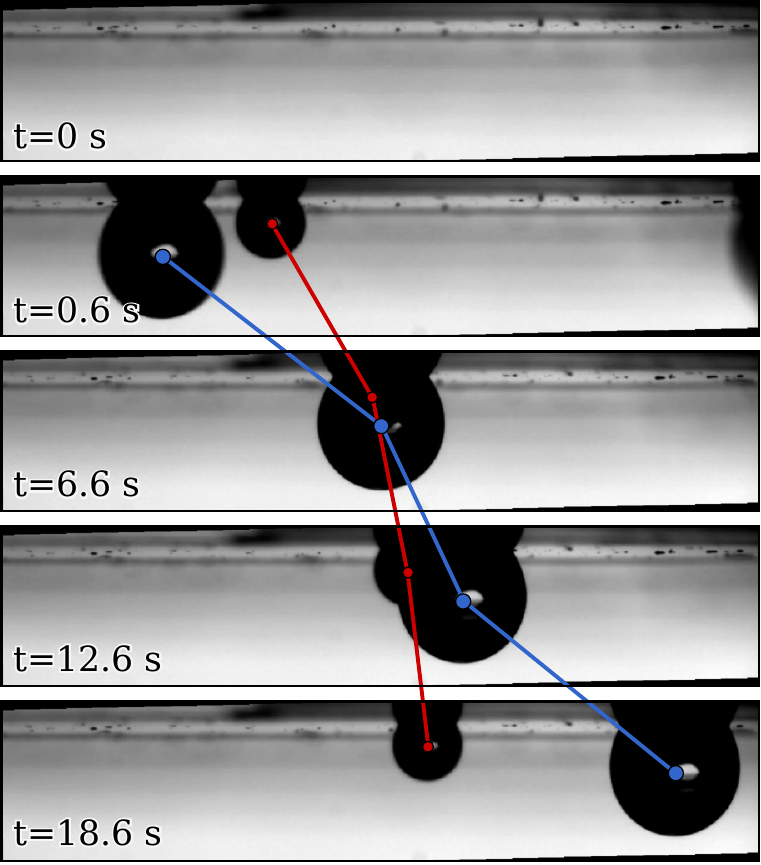}
  \caption{
    Droplets projected on an upside-down shaped liquid surface.
    Droplet are captured due to the surface high normal adhesion and then driven towards the high wettability region (right here).
    Droplets of different sizes move at different velocities, in agreement with Fig. 2 form the main manuscript.
  }
  \label{fig:hanging_droplets_1}
\end{figure}

\begin{figure}[H]
  \centering
  \includegraphics[width=.75\textwidth]{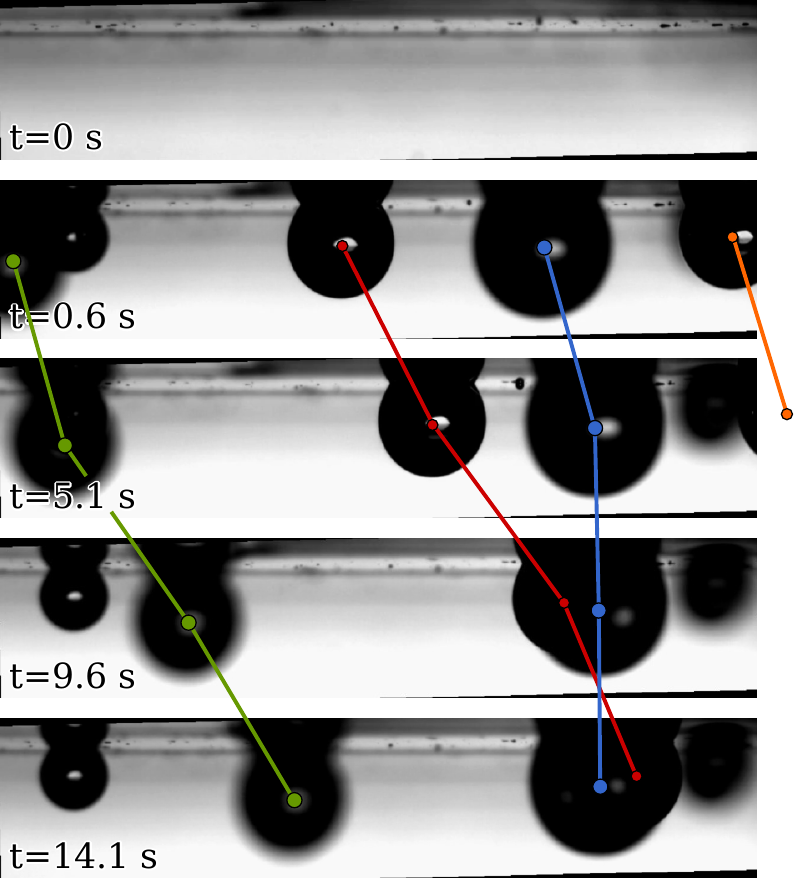}
  \caption{
    Same as Fig.~\ref{fig:hanging_droplets_1}, but for a higher number of droplets.
    Some droplets lands outside of the shaped region of the sample and remain motionless.
  }
  \label{fig:hanging_droplets_2}
\end{figure}

\textbf{We present here additional visualisations of hanging droplets moving due to wettability gradients (Figures \ref{fig:hanging_droplets_1} and \ref{fig:hanging_droplets_2}).}\\
(See also Movie S4).
The droplets are generated using a reservoir of water, whose free-surface is excited at its characteristic frequency using a shaker.
Droplets in a given range of size are successfully captured thanks to the high normal adhesion of the liquid surface.
Because of the inherent wettability gradient, droplets are then self-propelled towards the high wettability regions (right here).
Samples are carefully leveled prior to each experiments, to ensure that gravity is not playing any role in the droplets motion.

\subsection{Captions of the movies:}
\paragraph{Movie S1.}
Droplet rapid motion due to a discontinuity in wettability of a shaped liquid surface.
The solid fraction is $f_s=0.1$ on the left and $f_s=0.9$ on the right.
When released, the droplet moves rapidly towards the region of higher wettability (higher solid fraction).
The scale bar is \SI{1}{\mm}.

\paragraph{Movie S2.}
Droplet deposited on a shaped liquid surface with a gradient of solid fraction ($\alpha = \SI{0.05}{\per \mm}$).
The gradient of wettability and the low pinning force allow the droplet to move towards the high wettability regions.
The scale bar is \SI{1}{\mm}.

\paragraph{Movie S3.}
Top view of two droplets deposited on a shaped liquid surface with a gradient of solid fraction along a curved path.
Droplets move along the curved rails towards the region of higher wettability.
For convenience, movies are accelerated by a factor of 4.

\paragraph{Movie S4.}
Droplets projected on an upside-down shaped liquid surface.
Droplets are captured due to the surface high normal adhesion and then driven towards the high wettability region (right here).
Droplets of different sizes move at different velocities.
\end{document}